\nonstopmode

\newcommand{\E}[1]{\times 10^{#1}}
\newcommand{\im}{{\rm Im\,}}
\newcommand{\eV}{\U{eV}}
\newcommand{\degree}{{}^{\circ}}
\newcommand{\ie}{i.e.{}}
\newcommand{\eg}{e.g.{}}
\newcommand{\U}[1]{\,{\rm{#1}}}
\newcommand{\I}[1]{_{\mathrm{#1}}}
\newcommand{\imag}{{\rm i}}
\newcommand{\Sum}{\sum\limits}
\newcommand{\ket}[1]{\left|\right.\!#1\!\left.\right>}
\newcommand{\Sxray}{$\sigma_{1s, \mathrm{n.l.}}(\omega_{\mathrm{X}})$}
\newcommand{\Spar} {$\sigma_{1s}^{\parallel}   (\omega_{\mathrm{X}})$}
\newcommand{\Sperp}{$\sigma_{1s}^{\perp}       (\omega_{\mathrm{X}})$}
\renewcommand{\frac}[2]{{{#1}\over{#2}}}

\documentclass[prl,aps,twocolumn,showpacs,nopreprintnumbers]{revtex4}
\usepackage{bm,bbm,amssymb,graphicx}
\usepackage[nativepdf,bookmarks,bookmarksopen,bookmarksnumbered,raiselinks,%
pdftitle={Electromagnetically induced transparency for x rays},%
pdfauthor={Christian Buth, Robin Santra, Linda Young},%
pdfsubject={Atomic physics},%
pdfkeywords={quantum optics, ab initio, laser dressing, x rays, photoabsorption, cross section, %
polarization dependence, nonperturbative, laser-matter interaction, neon, %
electromagnetically induced transparency, EIT, pulse shaping, index of refraction}]{hyperref}

\begin{document}
\title{Electromagnetically induced transparency for x rays}
\author{Christian Buth}
\author{Robin Santra}
\author{Linda Young}
\affiliation{Argonne National Laboratory, Argonne, Illinois~60439, USA}
\date{10~May~2007}

\begin{abstract}
Electromagnetically induced transparency~(EIT) is predicted for x rays in laser-dressed
neon gas. The x-ray photoabsorption cross section and polarizability near the Ne {\em K} edge
are calculated using an \emph{ab initio} theory suitable for optical strong-field problems.
The laser wavelength is tuned close to the transition between $1s^{-1} \, 3s$ and
$1s^{-1} \, 3p$ ($\sim 800\,$nm). The minimum laser intensity required to observe EIT
is of the order of $10^{12}\,$W/cm$^2$. The \emph{ab initio} results are discussed in
terms of an exactly solvable three-level model. This work opens new opportunities for
research with ultrafast x-ray sources.
\end{abstract}

%
%
%

\preprint{arXiv:0705.3615}
\maketitle

\renewcommand{\onlinecite}[1]{\cite{#1}}

In a $\Lambda$-type medium characterized by atomic levels $a$, $b$, and $c$
with energies $E_a > E_b > E_c$, resonant
absorption on the $c \to a$ transition can be strongly suppressed
by simultaneously irradiating the medium with an intense laser that couples
the levels~$a$ and $b$.  This phenomenon is known as electromagnetically induced
transparency~(EIT)~\cite{Harris:NL-90,Boller:ET-91,Harris:EI-97}.
EIT enables one to control the absorption and dispersion of a gaseous medium.
It has become a versatile tool for creating
media with exceptional optical properties \cite{Hau:SL-99,PhysRevLett.86.783,%
Lukin:ET-01,Fleischhauer:ET-05}. EIT forms the basis of a recent proposal for
a high-accuracy optical clock \cite{Santra:HA-05}.

In this Letter, we study EIT for x rays. Specifically, we consider the near-{\em K}-edge
structure of neon gas in the presence of a linearly polarized $800$-nm laser with
an intensity of~$10^{13}\,$W/cm$^2$. The decay widths of excited states involved
in EIT in the optical regime typically do not exceed $\sim 10^{-4}\,$eV (see, \eg,
Ref.~\cite{Boller:ET-91}) and are often much smaller.  In contrast, the decay width of
core-excited neon, $\Gamma_{1s} = 0.27 \eV$~\cite{Schmidt:ES-97}, is larger by a factor
of~$\sim 2000$.  Therefore, as we will show below, the intensity of the
coupling laser must be extraordinarily high.  This represents the first case of EIT
where the laser causes strong-field ionization of the two upper levels.

Because of their high binding energy, the core and valence electrons of Ne remain essentially
unperturbed at~$10^{13}\,$W/cm$^2$. (Multiphoton ionization of Ne in its ground state is negligible.)
However, laser dressing of the core-excited states introduces strong-field physics: For the laser parameters
employed here, the ponderomotive potential~\cite{Freeman:PP-88} is
$U\I{p} = 0.60 \eV$, and the energy needed to ionize, for instance, the $3p$ electron in the
$1s^{-1} \, 3p$ state is $I_{1s^{-1}\,3p} = 2.85 \eV$  ($1s^{-1} \, 3p$ denotes the state
produced by exciting a $1s$ electron to the $3p$~Rydberg orbital). Hence, the Keldysh
parameter~\cite{Keldysh:-65} is $\gamma = \sqrt{I_{1s^{-1}\,3p}/(2 \, U\I{p})} = 1.5$.
For $\gamma \ll 1$, the atom--field interaction can be described in the adiabatic
tunneling picture; the perturbative multiphoton regime is indicated by~$\gamma \gg 1$.
In our case, where $\gamma \approx 1$, the nonadiabatic strong-field regime persists;
neither perturbation theory nor a tunneling description is adequate.
It should also be noted that the transition energy between, \eg, the states $1s^{-1}\,3s$
and $1s^{-1}\,3p$ is $1.69 \eV$, which is, within the decay width of the core-excited
states, in one-photon resonance with the 1.55-eV laser.

For this intermediate intensity regime, we have developed the \emph{ab initio} theory described
in Ref.~\cite{Buth:TX-up}.
Briefly, the electronic structure of the atom is described in the Hartree-Fock-Slater
approximation; nonrelativistic quantum electrodynamics is used to treat the
interaction of the atom with the two-color radiation field consisting of the dressing
laser and the x rays. Our calculations were carried out with the \textsc{dreyd}
program~\cite{fella:pgm-06}. We used the computational parameters from Ref.~\cite{Buth:TX-up},
with the exception that for the atomic basis, we included angular
momentum quantum numbers~$l = 0, \ldots, 9$, and for the laser-photon basis, we included
the simultaneous emission and absorption of up to 20~photons. The non-Hermitian Hamiltonian
matrix (dimension $41000$) was diagonalized using a Lanczos algorithm  for complex-symmetric
matrices \cite{0953-4075-31-18-009}. All results are converged with respect to the electronic
and photonic basis sets and the number of Lanczos iterations.

\begin{figure}
  \begin{center}
    \includegraphics[clip,width=\hsize]{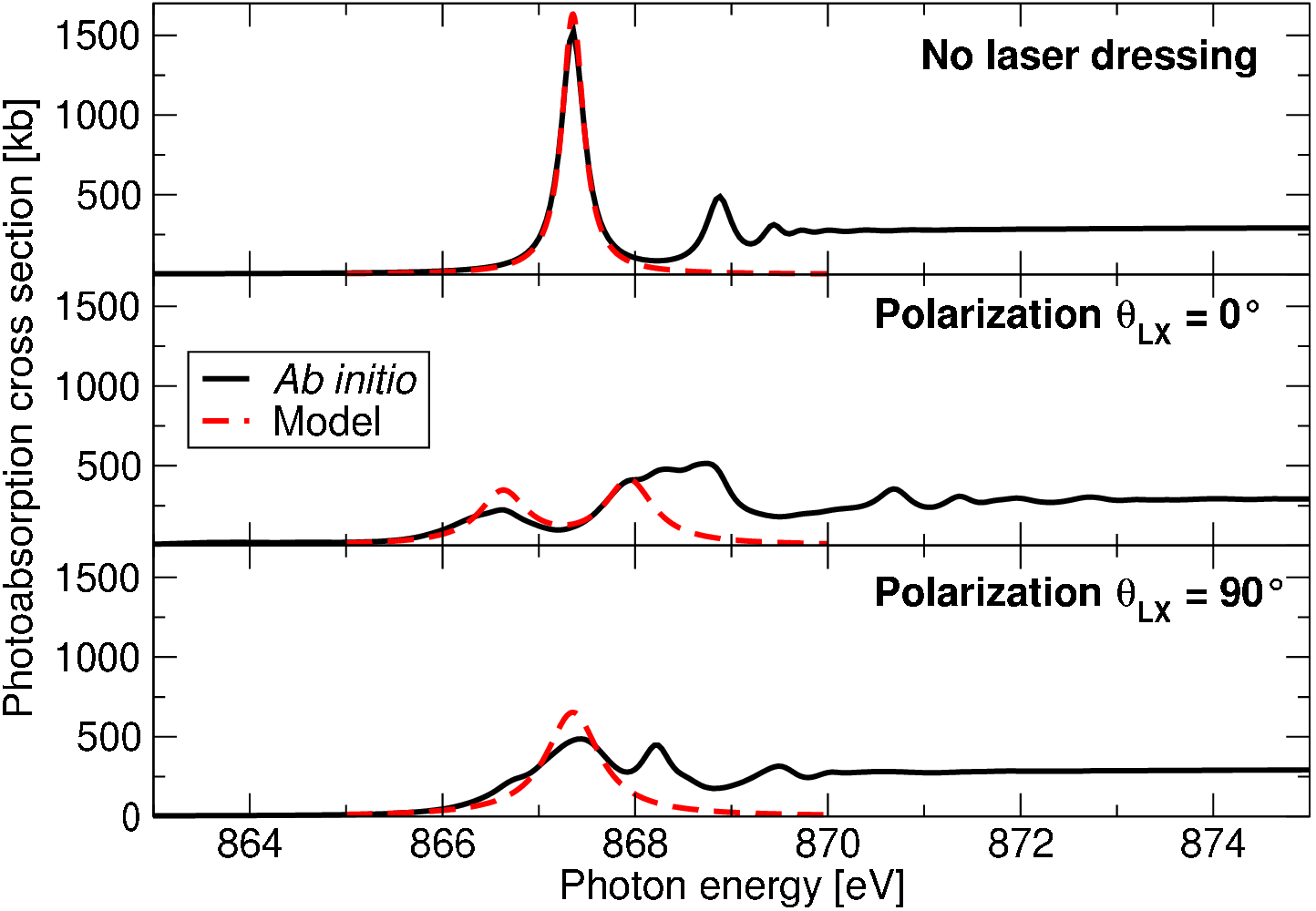}
    \caption{X-ray photoabsorption cross section of neon near the {\em K}~edge
             with laser dressing~[$\sigma_{1s}(\omega_{\mathrm X}, \theta_{\mathrm{LX}})$]
             and without it~[\Sxray{}].
             Results from \emph{ab initio} calculations and a three-level model are shown.
             Here, $\theta_{\mathrm{LX}}$ is the angle between the polarization vectors of
             the laser and the x rays. The laser operates at a wavelength of~$800 \U{nm}$
             and an intensity of~$10^{13}\,$W/cm$^2$.}
    \label{fig:NeI_3panels}
  \end{center}
\end{figure}

The calculated x-ray photoabsorption cross section of a neon atom is displayed in
Fig.~\ref{fig:NeI_3panels} for three cases.  (The x rays are assumed to be linearly
polarized.)  First, there is the cross section without laser dressing, \Sxray{}.
Second is shown the cross section for parallel laser and x-ray polarization
vectors, \Spar{}${}  \equiv \sigma_{1s}(\omega_{\mathrm X}, 0 \degree)$. In the
third case, the polarization vectors are perpendicular,
\Sperp{}${} \equiv \sigma_{1s}(\omega_{\mathrm X}, 90 \degree)$.
Our result for \Sxray{} is in good agreement with experimental and theoretical
data~\cite{Coreno:MA-99,Gorczyca:AD-00}. The prominent absorption feature
at~$867.4 \eV$ arises due to excitation into the $1s^{-1} \, 3p$ state. The weaker peaks
are associated with $1s \to np$ transitions with $n \ge 4$. The {\em K} edge lies at
$870.2 \eV$~\cite{Schmidt:ES-97}.

Exposing neon to intense laser light shifts and mixes the atomic energy levels. This leads to
the pronounced differences between the laser-on and laser-off cases in Fig.~\ref{fig:NeI_3panels}.
Nevertheless, in spite of the rather high laser intensity, there is still substantial structure
in both \Spar{} and \Sperp{}.  The absorption peaks are broader by about a factor of
two in comparison to the laser-free case. This is because the laser ionizes the Rydberg
electron at a rate that is comparable to the Auger decay rate of the {\em K}-shell hole. Therefore,
in the presence of the laser, one may expect an enhanced production of Ne$^{2+}$ in the pre-{\em K}-edge
region. The laser deforms the orbitals cylindrically along the laser polarization axis.
Thus, anisotropy arises with respect to the angle $\theta_{\mathrm{LX}}$ between x-ray and
laser polarization vectors. This is the origin of the differences between~\Spar{} and
\Sperp{}~\cite{Buth:TX-up}. The most eye-catching impact of the laser dressing can be seen in the
vicinity of the $1s \to 3p$ resonance, which is highly suppressed for
$\theta_{\mathrm{LX}} = 0^{\circ}$. As will be explained in the following paragraphs, this
suppression of the $1s \to 3p$ resonance is a manifestation of EIT. The suppression
for $\theta_{\mathrm{LX}} = 90^{\circ}$ is a consequence of laser-induced line broadening.
We would like to point out that, at least in the case of \Spar{}, weak
structures, not present in~\Sxray{}, emerge above the {\em K}~edge of the laser-free atom.
These are produced by multiphoton processes involving the photoelectron.
This is similar to above-threshold ionization~\cite{PhysRevLett.42.1127}.

In order to understand the reason for the laser-induced suppression of the $1s \to 3p$~transition,
we reduce the \emph{ab initio} theory of Ref.~\cite{Buth:TX-up} to a $\Lambda$-type model.
To this end, we identify the levels~$a$, $b$, and $c$ mentioned earlier with the states
$1s^{-1} \, 3p$, $1s^{-1} \, 3s$, and the atomic ground state, respectively. Levels~$a$ and $b$
are decaying states, due to both Auger decay and laser ionization, with widths~$\Gamma_a$
and $\Gamma_b$. The energy spacing between~$a$ and $b$ is symbolized by~$\omega_{ab}$;
similarly, $\omega_{ac}$ denotes the spacing between $a$ and $c$. The Hamiltonian $\hat H$
is the sum of the atomic Hamiltonian~$\hat H_{\mathrm{AT}}$, the free electromagnetic
field~$\hat H_{\mathrm{EM}}$, the laser--electron interaction~$\hat H_{\mathrm L}$, and the
x-ray--electron interaction~$\hat H_{\mathrm X}$. The Hamiltonian $\hat H$ is represented
in the direct product basis
$\{\ket{c}\ket{N_{\mathrm L}}\ket{N_{\mathrm X}},\ket{b}\ket{N_{\mathrm L}+1}\ket{N_{\mathrm X}-1},
\ket{a}\ket{N_{\mathrm L}}\ket{N_{\mathrm X}-1}\}$,
where $\ket{N_{\mathrm L}}$ and $\ket{N_{\mathrm X}}$ are Fock states of the laser and x-ray modes,
respectively. The strong laser coupling between $a$ and $b$ is treated by diagonalizing $\hat H$ in the
corresponding subspace. The x-ray coupling of level $c$ to the resulting dressed states is weak and
is treated perturbatively. The x-ray photoabsorption cross section obtained in this way is
\begin{widetext}
  \begin{equation}
    \label{eq:crosssection}
     \sigma_{\mathrm{Model}}(\omega_{\mathrm X}, \theta_{\mathrm{LX}})
      = \sigma_0 \ \im \Bigl[ \cos^2 \theta_{\mathrm{LX}} \Sum_{s = \pm} \Bigl(
        {\Omega_{ab}^2/4 \over E_s^2 + \Omega_{ab}^2/4}
        \; {\gamma_a \over \omega_{ac} - \imag \, \gamma_a + E_s - \omega_{\mathrm X}} \Bigr)
        + \sin^2 \theta_{\mathrm{LX}} \; {\gamma_a \over \omega_{ac}
        - \imag \, \gamma_a - \omega_{\mathrm X}} \Bigr] \; .
  \end{equation}
\end{widetext}
Here, $\Omega_{ab}$ is the Rabi frequency~\cite{Meystre:QO-91} associated with levels $a$ and $b$,
$E_{\pm} \equiv -\frac{1}{2} [\omega_{ab} - \omega_{\mathrm L} - \imag \,
(\gamma_a - \gamma_b)] \pm \, \frac{1}{2} \> \sqrt{(\omega_{ab} - \omega_{\mathrm L} - \imag \,
(\gamma_a - \gamma_b))^2 + \Omega_{ab}^2}$ is the level shift of the two laser-dressed states,
$\gamma_a = \Gamma_a/2$, $\gamma_b = \Gamma_b/2$, and $\omega_{\mathrm L}$ is the laser photon energy.
The parameter $\sigma_0$ is the cross section obtained on resonance ($\omega_{\mathrm X} = \omega_{ac}$)
without laser ($\Omega_{ab} \to 0$). Let the laser polarization axis coincide with the $z$ axis. Then,
at $\theta_{\mathrm{LX}} = 90^{\circ}$, the absorption cross section is given by a simple Lorentzian
[cf. Eq.~(\ref{eq:crosssection})] because only $1s^{-1} \, 3p_x$ or $1s^{-1} \, 3p_y$ can be excited
by the x rays. Dipole selection rules dictate that the laser can couple only $1s^{-1} \, 3p_z$ to
$1s^{-1} \, 3s$. The two dressed states that appear as a consequence of the laser-mediated coupling
can be probed by the x rays at $\theta_{\mathrm{LX}} = 0^{\circ}$. Using
$\Delta_{\mathrm{LX}} \equiv (\omega_{ac} - \omega_{\mathrm X}) - (\omega_{ab} - \omega_{\mathrm L})$,
we find from Eq.~(\ref{eq:crosssection}) for $\theta_{\mathrm{LX}} = 0^{\circ}$
\begin{widetext}
  \begin{equation}
   \label{eq:crosspara}
    \sigma_{\mathrm{Model}}(\omega_{\mathrm X}, 0^{\circ}) = \sigma_0 \;
        {\gamma_a^2 \, \Delta_{\mathrm{LX}}^2 + \gamma_a \, \gamma_b \, (\Omega_{ab}^2 / 4 + \gamma_a \, \gamma_b)
        \over [\Omega_{ab}^2 / 4 + \gamma_a \, \gamma_b
        - \Delta_{\mathrm{LX}} \, (\omega_{ac} - \omega_{\mathrm X})]^2
        + [\gamma_a \, \Delta_{\mathrm{LX}} + \gamma_b \, (\omega_{ac} - \omega_{\mathrm X})]^2} \; .
  \end{equation}
\end{widetext}
This expression is identical to the absorption cross section given in Ref.~\cite{Boller:ET-91}
for EIT in a $\Lambda$-type model. Because the lifetime of the laser-dressed core-excited
states considered here is only of the order of 1 fs, we neglect other decoherence mechanisms such as
collisional broadening. If the width $\gamma_b$ in Eq.~(\ref{eq:crosspara}) were zero
($\gamma_a \ne 0$, $\Omega_{ab} \ne 0$), then the absorption cross section would vanish completely
provided the resonance condition $\Delta_{\mathrm{LX}} = 0$ is satisfied. (See Ref.~\cite{Santra:HA-05} for an
example.) For finite $\gamma_b$, there is in general a suppression of absorption on resonance.
In view of Eq.~(\ref{eq:crosssection}), this suppression of absorption characteristic of EIT is due
to the destructive interference of the excitation pathways from the atomic ground state to the
two laser-dressed core-excited states.

The x-ray photoabsorption cross section of the model, Eq.~(\ref {eq:crosssection}), is displayed
in Fig.~\ref{fig:NeI_3panels} together with the {\em ab initio} data. The agreement for~\Sxray{}
is excellent, illustrating that the absorption line at~$867.4 \eV$ is nearly exclusively
caused by the $1s \to 3p$~transition. As far as level energies and transition dipole matrix elements
are concerned, the model and {\em ab initio} calculations share the same input data. However, in
the {\em ab initio} treatment, the laser is allowed to couple $1s^{-1} \, 3s$ and $1s^{-1} \, 3p$
to states outside the three-level subspace. This coupling leads to both additional AC Stark shifts~\cite{Meystre:QO-91}
and ionization broadening of levels $a$ and $b$. We account for the ionization broadening in the
presence of the laser by assuming effective linewidths~$\Gamma_a = 0.675 \eV$ and
$\Gamma_b = 0.54 \eV$. These parameters give reasonable agreement with the {\em ab initio}
results for \Spar{} and \Sperp{} in the vicinity of the $1s \to 3p$ resonance.  The AC Stark shifts
due to states outside the three-level subspace are apparently small and are neglected in the model.
The model reproduces the double-hump structure in \Spar{} and, in particular, the suppression of
the $1s \to 3p$ resonance. This agreement leads us to conclude that the dominant physics here is EIT.

Because of the rather large decay widths $\Gamma_a$ and $\Gamma_b$, the EIT structure seen in
\Spar{} is relatively insensitive to changes $\Delta\omega_{\mathrm L}$ of the laser
photon energy. We find that EIT persists for $\Delta\omega_{\mathrm L}/\omega_{\mathrm L} \approx 30 \%$.
In accordance with the approximate EIT resonance condition $\Delta_{\mathrm{LX}} = 0$, the x-ray
absorption minimum is shifted by $\sim \Delta\omega_{\mathrm L}$. A simple criterion for the laser
intensity required to observe a significant EIT effect can be found as follows.
The laser photon energy $\omega_{\mathrm L}$ is chosen to be in exact resonance with levels $a$
and $b$. EIT appears when the real parts of the two pole positions in Eq.~(\ref{eq:crosssection})
[$\theta_{\mathrm{LX}} = 0^{\circ}$] are separated by more than the Auger width~$\Gamma_{1s}$
(laser broadening of the lines is neglected here). The required laser intensity thus follows from
the relation~$|\Omega_{ab}| > \Gamma_{1s}$; we calculate~$I\I{L} > 4.3 \E{11}\,$W/cm$^2$.
This estimate is, in fact, somewhat too optimistic: We varied the intensity in the \emph{ab initio}
calculation for an 800-nm dressing laser and found appreciable EIT
for~$I\I{L} > 10^{12}\,$W/cm$^2$. The EIT signature is particularly clear for intensities
around~$10^{13}\,$W/cm$^2$, the value used for Fig.~\ref{fig:NeI_3panels}. Because of the strong
perturbation by the laser field, the splitting between the two EIT peaks is only approximately
given by the Rabi frequency $|\Omega_{ab}|$.

\begin{figure}
  \begin{center}
    \includegraphics[clip,width=\hsize]{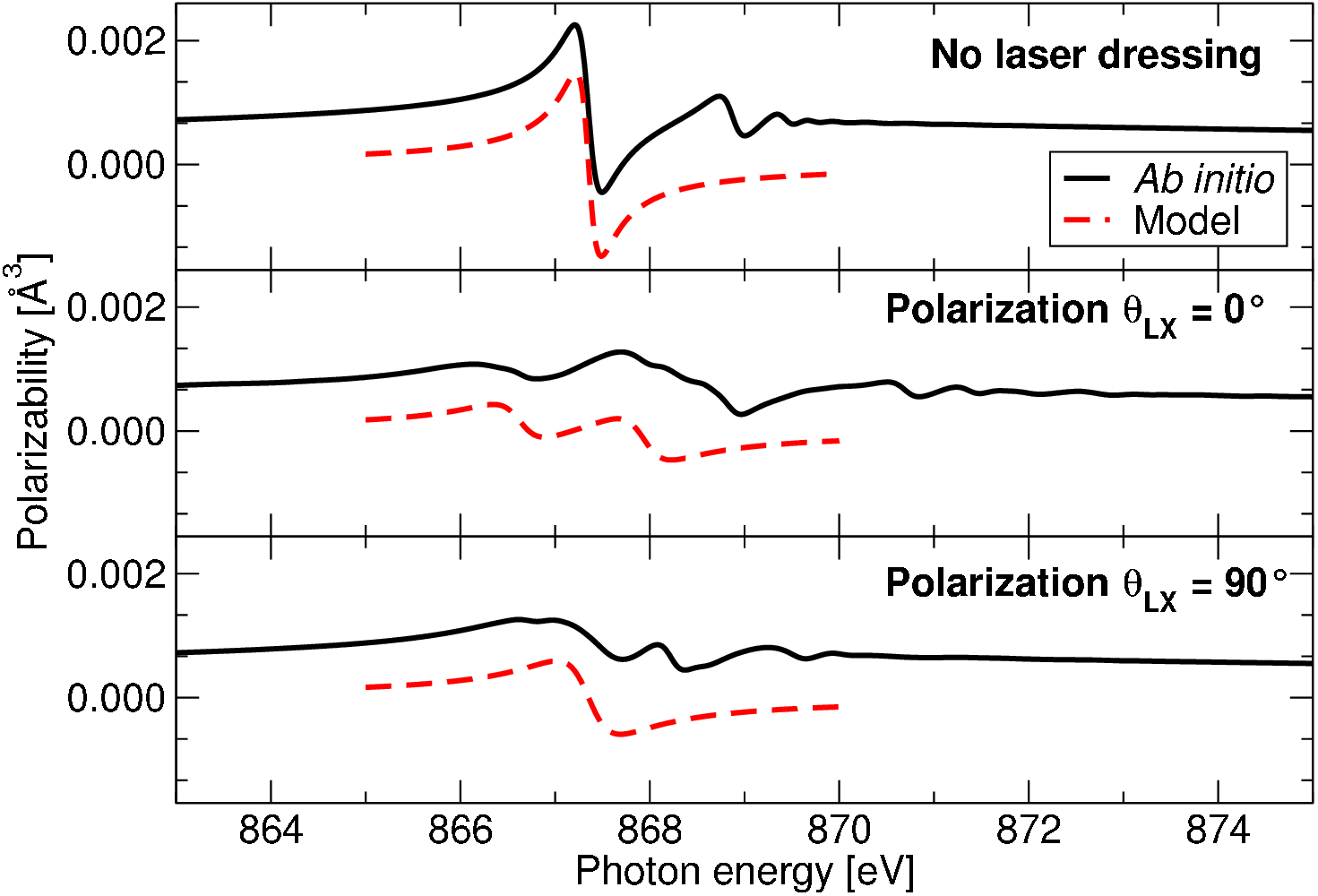}
    \caption{Dynamic polarizability of atomic neon near the {\em K} edge.
             See the caption of Fig.~\ref{fig:NeI_3panels} for details.}
    \label{fig:polarizability}
  \end{center}
\end{figure}

For a complete characterization of the x-ray properties of laser-dressed atoms, we need---in addition
to the absorption cross section---the dynamic (\ie{}, photon-energy-dependent) polarizability
$\alpha(\omega_{\mathrm X},\theta_{\mathrm{LX}})$. As will be described elsewhere, $\alpha(\omega_{\mathrm X},\theta_{\mathrm{LX}})$
may be calculated using an extension of the formalism of Refs.~\cite{Buth:TX-up} and \cite{Buth:NH-04}.
The dynamic polarizability of neon is plotted in Fig.~\ref{fig:polarizability} for the
three cases considered in Fig.~\ref{fig:NeI_3panels}. Both \emph{ab initio} and three-level model results are depicted.
Apart from a missing background due to states outside the three-level subspace, the model reproduces the
structure of $\alpha(\omega_{\mathrm X},\theta_{\mathrm{LX}})$ in the vicinity of the $1s \to 3p$~resonance.
The dynamic polarizability (together with the number density and the number of electrons per atom) determines the index of refraction.
As one can see in Fig.~\ref{fig:polarizability}, the primary impact of laser dressing on $\alpha(\omega_{\mathrm X},\theta_{\mathrm{LX}})$
is the suppression of dispersion near the resonance. This is in contrast to the situation encountered in the optical domain
\cite{Fleischhauer:ET-05}. Using the \emph{ab initio} data shown in Fig.~\ref{fig:polarizability}, we calculate that in the EIT
case ($\theta_{\mathrm{LX}} = 0^{\circ}$), the x-ray group velocity at the $1s \to 3p$~resonance is reduced relative to the speed of light
by only $3\times10^{-6}$. (A neon pressure of one atmosphere is assumed.) It therefore seems likely that EIT is most suitable for amplitude
modulation, rather than phase modulation, of x rays.

\begin{figure}
  \begin{center}
    \includegraphics[clip,width=\hsize]{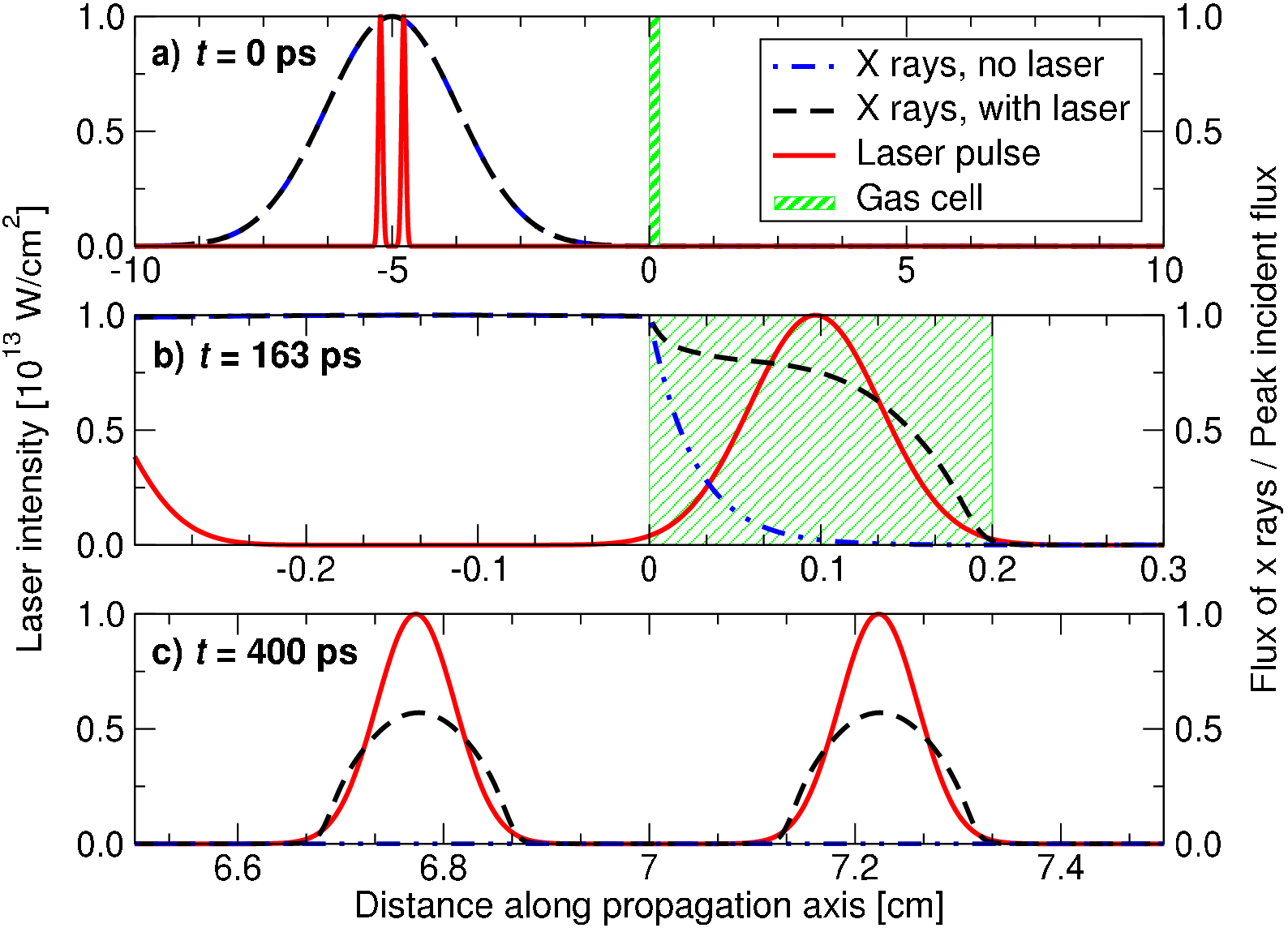}
    \caption{EIT-based generation of two ultrashort x-ray pulses with well-defined time delay.}
    \label{fig:transmission}
  \end{center}
\end{figure}

In practice, high intensities are achieved using short-pulse lasers. For instance, if one operates a Ti:Sapphire
laser system at a repetition rate of $\sim 1\,$kHz, the typical pulse energy per shot is $\sim 1\,$mJ. With a
focal width of, say, $50\,\mu$m, the pulse duration required for a peak intensity of $10^{13}\,$W/cm$^2$
is of the order of $1\,$ps. In order to {\em maximize} the EIT signal, the laser intensity during the exposure of the
laser-dressed medium to the x rays must remain high and essentially constant, i.e., the x rays must
have a pulse duration of $1\,$ps or shorter. Such ultrashort x-ray pulses can presently be produced by high-order
harmonic generation~\cite{Spielmann:GC-97}; laser plasma sources~\cite{RosePetruck:LP-03}; and the laser slicing
method~\cite{Schoenlein:FS-00,Khan:FS-06}. All three approaches provide a relatively low number of x-ray photons
per pulse. The upcoming x-ray free-electron lasers~\cite{LCLS:CDR-02,Tanaka:SCSS-05,Altarelli:TDR-06}
will be high-flux sources of femtosecond x-ray pulses.

A possible experimental scheme for observing the EIT effect discussed here is the following.
Consider a 2-mm-long gas cell filled with neon at atmospheric pressure and room temperature,
so that the neon number density is $2.4\times 10^{19}\,$cm$^{-3}$. In the absence of the laser,
the x-ray absorption cross section on the $1s \to 3p$ resonance is $\sim 1500\,$kb. Therefore,
the linear absorption coefficient is $36\,$cm$^{-1}$. Assuming that Beer's law
\cite{Meystre:QO-91} holds, only $0.07 \%$ of the x rays are transmitted through
the gas cell on resonance. Applying the dressing laser with parallel laser and x-ray polarizations,
the x-ray absorption cross section is suppressed by a factor of~$\sim 13$ (see Fig.~\ref{fig:NeI_3panels}).
In this case, the linear absorption coefficient is $2.8\,$cm$^{-1}$, and the x-ray transmission on
resonance rises to $57 \%$---an increase by a factor of $\sim 800$. This effect should be
straightforward to measure experimentally.

An exciting prospect is to use this scheme to imprint pulse shapes of the optical dressing laser onto the x rays.
To demonstrate the idea, we have numerically propagated a $100$-ps x-ray pulse (peak flux $8\times10^9$ photons/ps/cm$^2$)
together with two $3$-ps laser pulses, taking into account the laser intensity dependence of the $1s \to 3p$
x-ray absorption cross section in the neon gas cell.  Three temporal snapshots are shown in Fig.~\ref{fig:transmission}.
At $t=0$, the x-ray and laser pulses are still outside the gas cell. After $163\,$ps, the first of the two laser pulses
overlaps with the x-ray pulse inside the gas cell, thereby substantially enhancing x-ray transmission in comparison to
the laser-off case. At $t=400\,$ps, i.e., after propagation through the gas cell, two ultrashort x-ray pulses ($\sim 5\,$ps)
emerge. The x-ray pulses are somewhat broadened with respect to the laser pulses because of the nonlinear
dependence of the x-ray absorption cross section on the laser intensity. The time delay between the two x-ray pulses
can be controlled by changing the time delay between the two laser pulses, opening a route to ultrafast all x-ray
pump-probe experiments. With an analogous strategy, controlled shaping of short-wavelength pulses might become a reality,
similar to the sophisticated techniques already available for optical wavelengths~\cite{Weiner:FS-00}.
Estimates show that for the parameters quoted, dispersion effects due to neutral neon atoms dominate over those caused
by the rather dilute x-ray-generated plasma. The time delay experienced by both x-ray and laser pulses transmitted through
the gas cell is in the sub-fs regime.

\begin{acknowledgments}
We thank T.~E.~Glover and A. Wagner for fruitful discussions.
This work was supported by the Office of Basic Energy Sciences, Office of Science,
U.S.~Department of Energy, under Contract No.~DE-AC02-06CH11357.
C.B.~was funded by the Alexander von Humboldt Foundation.
\end{acknowledgments}

\end{document}